\documentclass[12pt]{article}

\usepackage{siunitx}
\usepackage{graphicx}
\usepackage{scicite}
\usepackage{times}
\usepackage{float}
\usepackage{multibib}
\usepackage{amsmath}
\usepackage{amssymb}
\usepackage{physics}

\newenvironment{sciabstract}{%
\begin{quote} \bf}
{\end{quote}}

\title{Encapsulated void resonators in lossy dielectric van der Waals heterostructures}

\author{
Avishek Sarbajna,$^{1,\dagger}$ Dorte Rub{\ae}k Danielsen,$^{1,\dagger}$ Laura Nevenka Casses,$^{2,3}$ \\ Nicolas Stenger,$^{2,3}$ Peter B{\o}ggild,$^{1}$ S{\o}ren Raza$^{1,\ast}$\\
\\
\normalsize{$^{1}$Department of Physics, Technical University of Denmark,}\\
\normalsize{Fysikvej, DK-2800 Kongens Lyngby, Denmark.}\\
\normalsize{$^{2}$Department of Electrical and Photonics Engineering, Technical University of Denmark,}\\
\normalsize{{\O}rsteds Plads, DK-2800 Kongens Lyngby, Denmark.}\\
\normalsize{$^{3}$NanoPhoton, Center for Nanophotonics, }\\
\normalsize{{\O}rsteds Plads, DK-2800 Kongens Lyngby, Denmark.}\\
\\
\normalsize{$^\dagger$These authors contributed equally to this work.} \\
\normalsize{$^\ast$Corresponding author: S{\o}ren Raza, E-mail: sraz@dtu.dk}
}

\date{}

\begin{document} 


\baselineskip24pt

\maketitle


\begin{sciabstract}
Dielectric optical resonators traditionally rely on materials with the combination of high refractive indices and low optical losses. Such materials are scarce for operation in visible spectrum and shorter wavelengths. This limitation can be circumvented by relaxing the requirement of low losses. We demonstrate that highly lossy dielectric materials can be structured to support optical resonances that confine light in air voids. We theoretically design void resonances in the visible spectrum and identify resonant modes supported by void arrays. Experimentally, we fabricate void arrays in tungsten diselenide and characterize the confined resonances using far-field reflectance measurements and scanning near-field optical microscopy. Using van der Waals heterostructure assembly, we encapsulate the voids with hexagonal boron nitride which reduces the void volume causing a large spectral blue shift of the void resonance exceeding 150~nm. Our work demonstrates a versatile optical platform for lossy materials, expanding the range of suitable materials and the spectral range of photonic devices.

\end{sciabstract}

\section*{Teaser}
Demonstration of resonant confinement in void arrays and post-fabrication resonance tuning by van der Waals encapsulation.


\section*{Introduction}
Controlling light-matter interaction at the nanoscale is a fundamental pursuit of nanophotonics, finding applications in biological sensing, structural colouration, imaging, and quantum technologies~\cite{novotny2011antennas, Kristensen2016, Tommasi:2021, Santiago-Cruz2022, Yang2023}. Achieving this control is primarily accomplished through the enhancement of the amplitude of the light field via optical resonances in nanostructured optical materials. Among these, dielectric materials with high refractive indices have stood out due to their ability to enhance both the electric and magnetic components of light~\cite{Kuznetsov:2016}. These nanostructures, known as dielectric Mie resonators, confine light within the dielectric material~\cite{Kruk2017} and have facilitated breakthroughs in nonlinear optics~\cite{Liu2016}, entangled photon generation~\cite{Marino2019}, enhanced Raman scattering~\cite{Raza2021}, directional scattering~\cite{Fu2013} and emission~\cite{Cihan2018}, underscoring the significant impact of Mie resonators in the advancement of nanoscale optics. Despite these achievements, Mie resonators are constrained by their requirement on near-lossless materials to ensure that the resonantly enhanced light is reradiated and not absorbed. This means that operation must occur below the band gap energy of the material, which not only narrows the scope of suitable materials but also restricts access to higher energy spectral domains, notably the blue and ultraviolet domains~\cite{Baranov2017}. Although, recent efforts have been made to discover new dielectric materials with wider band gaps~\cite{Svendsen2022,Doiron2022,Khurgin2022}, most Mie resonators are based on a narrow range of semiconductor materials, including silicon~\cite{Fu2013,Assadillayev:2021}, III-V materials~\cite{Liu2016,Cambiasso2017,Jang2020}, and, more recently, transition metal dichalcogenides (TMDCs)~\cite{Verre:2019}.

The complementary geometry where light is confined within air surrounded by a high-refractive-index material, so-called Mie voids, is not subject to the same material constraints. Mie voids have been theoretically studied~\cite{Chen:1998,Hamidi:2023} and recently demonstrated experimentally for the first time in silicon~\cite{Hentschel:2023}. By confining light in air, these resonators can be operated with optical materials that absorb light, thus allowing for functionality above the band gap energy. This development would diversify the semiconductor materials that can be employed, by including those with lossy characteristics traditionally considered incompatible with resonator designs. Operation above the band gap energy also naturally extends the accessible spectral range to higher frequencies, opening up the possibility of constructing metasurfaces for ultraviolet lithography~\cite{Zhao2021} or matching the natural electronic transitions of biomolecules for label-free molecular sensing~\cite{Chowdhury2009} and circular dichroism spectroscopy~\cite{Hu2020}.

\begin{figure}[H]
    \centering
    \includegraphics[width=0.5\columnwidth]{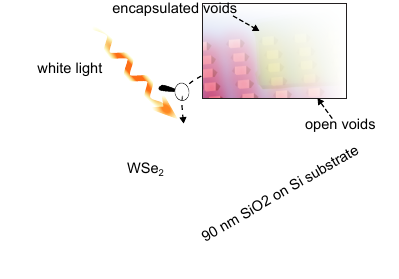}
    \caption{\textbf{Schematic of encapsulated void resonators.} Arrays of nanostructured holes in lossy bulk WSe$_2$ can support resonant modes confined to the void regions (red colour). Encapsulating the voids with a thin van der Waals material reduces the void volume, which gives rise to a blueshift of the resonance and thereby a colour change.}
    \label{fig_1}
\end{figure}

Here, we experimentally realise void resonators in bulk tungsten diselenide (WSe$_2$), operated above its band gap energy ($\sim 1.2$~eV). To aid in our design, we simulate the optical response of a system, consisting of an array of cylindrical voids in optically thick WSe$_2$, and separate different resonant contributions. Our simulations and the presence of void modes are supported by both far-field reflectance measurements and scanning near-field optical microscopy. Taking advantage of van der Waals heterostructure assembly, we transfer thin flakes of both WSe$_2$ and hexagonal boron nitride (hBN) to encapsulate the voids and thereby add another degree of freedom to control the void resonance (see Fig.~\ref{fig_1}). We find that a thin WSe$_2$ ($\sim9$~nm) or a thicker hBN ($\sim55$~nm) encapsulation allows us to engineer the void volume and shift the void resonance to shorter wavelengths, with hBN encapsulation ensuring that light couples in unimpeded due to its transparency in the visible range. With the increasing availability of van der Waals materials~\cite{Zotev2023}, our architecture offers a flexible platform, for expanding the choice of materials and spectral range of operation of nanoscale optical resonators. 

\section*{Results} 
\subsection*{Theory} 
\begin{figure}[H]
    \includegraphics[width=\textwidth]{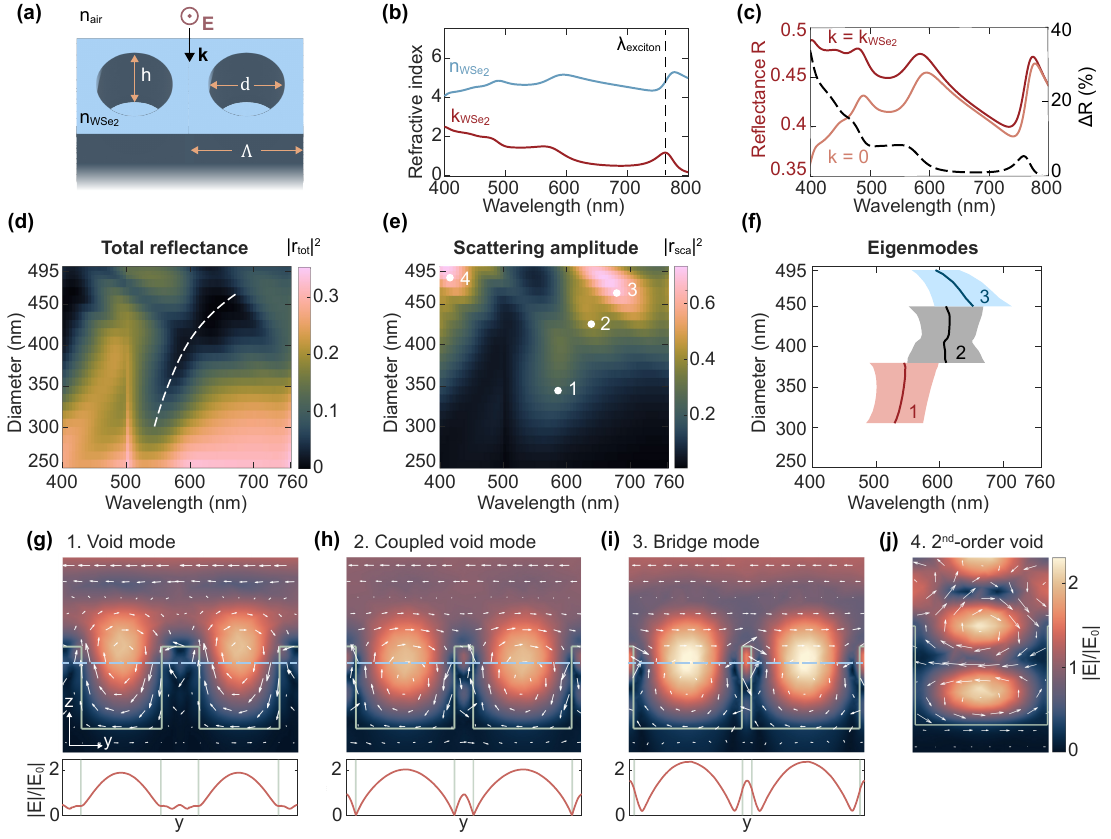}
    \caption{\textbf{Formation and coupling between void resonances in WSe$_2$.} \textbf{(a)} Schematic illustration of WSe$_2$ voids with depth $h$, diameter $d$, and period $\Lambda$. \textbf{(b)} In-plane real (blue) and imaginary (red) part of the refractive index of WSe$_2$. Data is from Ref.~\cite{Munkhbat:2022}.
    \textbf{(c)} Left axis: Reflectance of flat semi-infinite WSe$_2$ (red solid) and a material with $(n, k) = (n_{\textrm{WSe}_2}, 0)$ (light red solid). Right axis: Percentwise increase in reflectance when $k = k_{\textrm{WSe}_2}$ compared to $k = 0$ (black dashed). \textbf{(d,e)} Simulated total reflectance (d) and scattering amplitude (e) from a system of voids with $h = \SI{350}{nm}$, $\Lambda = \SI{500}{nm}$, and increasing diameter. Four resonances are visible in (e). The resonances are classified as (1) the void mode, (2) coupled void mode, (3) the bridge mode, and (4) second-order void mode. \textbf{(f)} Eigenfrequencies of the first three modes as a function of void diameter. Solid lines (shaded regions) show the real (imaginary) part of the eigenvalues. \textbf{(g-j)} Electric (colour map) and magnetic (white arrows) field distributions from the four regions circled in (e). Bottom panels, normalized electric field along line cuts outlined by the blue dashed lines.}
    \label{fig_2}
\end{figure}

We consider a periodic array (period $\Lambda$) of cylindrical voids with diameter $d$ and depth $h$ fabricated in thick WSe\textsubscript{2} flake~(see Fig.~\ref{fig_2}a). Tungsten diselenide exhibits both a high in-plane refractive index $n_{\textrm{WSe}_2}$ and a high in-plane extinction coefficient $k_{\textrm{WSe}_2}$ in the visible range~(see Fig. \ref{fig_2}b). Both of these material properties are beneficial for the formation of void modes as they increase the reflectance of the cavity walls. 
To illustrate the impact of material loss on reflectance, Fig.~\ref{fig_2}(c) shows the reflectance of bulk WSe$_2$ (solid red line) of normally incident light calculated using Fresnel's equation \cite{Born:1999}, and the reflectance of WSe$_2$ where the extinction coefficient is artificially set to zero (solid light red line). The reflectance scales with increasing extinction coefficient, leading to a large percentage-wise increase in the low wavelength region, with up to \SI{33}{\%} at \SI{400}{nm} (dashed black line). 

To understand the interaction between light and the void arrays, we simulate the zeroth-order reflectance of void arrays embedded in a WSe$_2$ substrate. The simulations are performed in the commercial finite-element-solver COMSOL Multiphysics v6.1~(see Methods). The void arrays have period $\Lambda = \SI{500}{nm}$, depth $h = \SI{350}{nm}$, and varying diameter $d = 250-\SI{495}{nm}$ (Fig.~\ref{fig_2}a). The chosen array period ensures that diffraction effects remain outside of the spectral regime of the fundamental void modes. The simulated reflectance spectra show a reflectance dip at a wavelength above the void period that shifts to longer wavelengths as the diameter increases~(Fig.~\ref{fig_2}d). However, as the diameter increases above approximately 400~nm additional reflectance minima appear, which indicate an interplay between multiple resonances.
To understand this, we subtract the reflection of a flat WSe$_2$-air interface from the total reflection coefficient to determine the amplitude of the scattered light produced by the voids. The scattering amplitude is linked to the resonant pathways in our system. We decompose the total reflection coefficient ($r_\textrm{tot}$) as
\begin{equation}
r_\mathrm{tot} = r_{\mathrm{WSe}_2} + r_\mathrm{sca},
\label{eq:r_mie}
\end{equation}
where $r_{\textrm{WSe}_2}=(1-\tilde{n}_{\textrm{WSe}_2})/(1+\tilde{n}_{\textrm{WSe}_2})$ is the reflection coefficient of the flat WSe$_2$-air interface and $r_{\textnormal{sca}}$ denotes the contribution due to the scattered light. Here, $\tilde{n}_{\textrm{WSe}_2}=n_{\textrm{WSe}_2}+i k_{\textrm{WSe}_2}$ is the complex in-plane refractive index of WSe$_2$. The diameter-independent reflection coefficient $r_{\textrm{WSe}_2}$ describes the background field set up in the absence of the voids. Thus, this procedure is similar to the scattered field formulation typically applied to individual scatterers, but here it is used for our periodic system to determine the scattered light due to the presence of the voids. 

The amplitude of the scattering coefficient $\abs{r_{\textrm{sca}}}^2$ shows three maxima for wavelengths larger than the period ($\lambda>500$~nm) that are present at different void diameters and also undergo different spectral shifts as the diameter increases~(Fig.~\ref{fig_2}e). As the resonant pathways are described by Lorentzian contributions to the reflection coefficient, these three maxima indicate the contribution of three eigenmodes in the void array. To confirm this, we perform eigenmode simulations (see Methods) and track both the real and imaginary parts of the eigenfrequency as the diameter increases~(Fig.~\ref{fig_2}f). These simulations confirm the presence of three eigenmodes with spectral shifts and linewidth changes, given by the real and imaginary part of the eigenfrequency, respectively, that depend on the void diameter. For small void diameters, the first eigenmode redshifts as the diameter increases due to the increasing void volume, consistent with the behaviour of individual Mie void resonators~\cite{Hentschel:2023}. Field plots of the first eigenmode show that the electric and magnetic field are localized within the void, with only a small component inside the WSe$_2$ bridge (Fig.~\ref{fig_2}g). The nature of this mode is reminiscent of the Kronig--Penney model in quantum mechanics consisting of a periodic array of rectangular potential barriers. Tungsten diselenide is very lossy in the spectral range of the first eigenmode (see Fig.~\ref{fig_2}b), which strongly dampens the propagating fields inside the WSe$_2$ bridges and therefore acts as a barrier. Consequently, the voids are nearly uncoupled and we denote this eigenmode as a void mode.

The second eigenmode appears in the diameter range $380-450$~nm with a varying linewidth but nearly constant resonance wavelength. Examining the field plots shows that the electromagnetic fields in the voids are similar to the void mode discussed previously, but that the field structure inside the WSe$_2$ bridge has changed (Fig.~\ref{fig_2}h). Whereas the void mode contains four field nodes and a small field amplitude in the bridge region (Fig.~\ref{fig_2}g), the second eigenmode shows a larger amplitude in the bridge with only two nodes. The increase in the field amplitude is due to a decreased extinction coefficient of WSe$_2$ in the spectral range of the second eigenmode. This leads to a coupling between neighboring voids and competing mechanisms for the resonance wavelength. When the diameter increases, the the void volume increases as well, while the bridge width decreases. These are mechanisms that individually should induce a redshift or blueshift of the resonance wavelength, respectively. The electromagnetic field of the second eigenmode is distributed such that these competing mechanisms nearly cancel out and give rise to the near-constant resonance wavelength across a wide range of void diameters. We refer to this eigenmode as a coupled void mode.

The third eigenmode is present for diameters larger than $450$~nm and blueshifts as the diameter increases (Fig.~\ref{fig_2}f). This eigenmode is in a wavelength regime where the extinction coefficient and thereby the optical loss decreases even further, giving rise to a significant field amplitude in the bridge region (Fig.~\ref{fig_2}i). The width of the bridge region therefore dictates the position of the resonance wavelength, leading to a blueshift as the diameter increases, and we refer to this eigenmode as the bridge mode.
Finally, the field plot in Fig.~\ref{fig_2}j shows that the fourth branch (labelled 4 in Fig.~\ref{fig_2}e) in the wavelength range below the period ($\lambda<500$~nm) stems from the second-order void mode.

\subsection*{Void resonances in a partially open system}
We experimentally verify the void resonances by fabricating voids with diameters spanning from 275-425$\,$nm and corresponding depths between 295-380$\,$nm on 438$\,$nm and 422$\,$nm thick mechanically exfoliated flakes (see Methods for fabrication procedure). The flakes are not completely etched through, leaving a considerably thick residual WSe$_2$ layer beneath the voids to avoid substrate effects (see Section~S1 and Fig.~\ref{fig:FlakeThicknessSweep_and_SiO2SiSubsComp_SI}). Simulations and reflection measurements of the samples with different residual WSe$_2$ thicknesses show that the thickness of these layers have minimum effect on the spectral features of the voids (Fig.~\ref{fig:FlakeThicknessSweep_and_SiO2SiSubsComp_SI}, Fig.~\ref{fig:Reflection figs}, and Section~S2). This ensures that we can compare the optical responses of the voids fabricated on flakes with different thicknesses. Voids of the same diameters and depths are arranged in a $30\times30$ array to ensure a sufficiently large signal in far-field reflection measurements. As the hole diameter influences the etch rate in WSe$_2$ \cite{Danielsen:2021}, the hole depth increases with diameter (see Section~S3 and Fig.~\ref{fig:Diam_vs_depth}). In the following, any reference to voids of different diameters also implies voids of different depths.

Bright-field (BF) imaging reveals that voids of different diameters exhibit bright changing colours within the visible spectrum (Fig.~\ref{fig_3}a). The corresponding scanning electron microscopy (SEM) images provide a detailed visualisation of the fabricated structures. The hexagonal morphology of the voids arises due to the anisotropic etching process employed during the fabrication~\cite{Danielsen:2021}. Since the simulated reflectance spectra of circular and hexagonal voids are close to identical (see Section~S4 and Fig.~\ref{fig:SharpPeak_test_SI}), we define an effective diameter of the voids as $d = \sqrt{4A/\pi}$, where $A$ is the cross-sectional area of the hexagonal hole measured by SEM (Section~S5 and Fig.~\ref{fig:Diam_SEM}). The void depths are measured by atomic force microscopy (AFM)~(Fig.~\ref{fig:topography AFM}).

Figure~\ref{fig_3}b shows the experimental (solid) and simulated (dashed) reflectance spectra from voids with different diameters and depths. We normalise the reflected signal from the voids to the signal from a flat region of the flake to eliminate the contribution from the bare WSe$_2$ flake (see Methods). The experimental and simulated spectra exhibit broad reflectance dips. For the voids with $d = \SI{340}{nm}$ and $d = \SI{360}{nm}$, the broad dips are accompanied by discernible shoulders indicating signs of multiple resonances. As the diameter increases to \SI{390}{nm} and \SI{425}{nm}, the reflectance dips narrow. For these void diameters, experimental results also show reflectance dips at shorter wavelengths (around \SI{430}{nm}), which indicates the presence of the second-order void mode, further corroborating  the agreement between experimental findings and theoretical predictions. 

The scattering amplitude of the resonant pathways $\abs{r_{\mathrm{sca}}}^2$ of the voids also demonstrates the involvement of multiple modes~(Fig.~\ref{fig_3}c). For the smallest void array ($d = \SI{275}{nm}$) the scattering amplitude shows that the reflectance dip at the wavelength of 580~nm corresponds to the void mode (red line in Fig.~\ref{fig_3}c).
The spectrally broad nature of the void mode suggests that these voids are at the onset of the void mode formation where the mode is not fully confined, leading to broadening due to radiation damping. Despite the lack of confinement, the single peak without splitting in the scattering amplitude is a sign that the coupling between neighbouring voids is minimal. With increasing void diameter the coupled void mode and bridge mode appear as depicted by the grey and blue lines in Fig.~\ref{fig_3}c, respectively. The resonance wavelength of the coupled mode has only a weak dependence on the void diameter, while the bridge mode blueshifts. These trends are in full agreement with the results shown in Fig.~\ref{fig_2}. For the largest void diameter ($d=425$~nm), the bridge mode is dominant which leads to a narrower reflectance dip both in the simulations and the experiments. 

Because of the WSe$_2$ etching process, the side-walls of the voids are not vertical but exhibit a sizable undercut. As a result, the larger holes with diameters $d = 390$~nm and $d=425$~nm have merged, forming a small hole in the upper part of the WSe$_2$ bridge (see tilted SEM images in Fig.~\ref{fig_3}d-e), while the smaller voids are fully separated (see Fig.~\ref{fig_3}f). Although merging is likely to enhance the coupling between the voids, the same trends that we see for non-merged voids continue for the merged voids. The exact shape of the undercut is difficult to measure. However, 2D simulations show good agreement between the void resonances in voids with an arbitrarily chosen undercut and voids of the same volume without an undercut, even when the voids are merged~(see Section~S6, Fig.~\ref{fig:ucut_sim_geometry}, and Fig.~\ref{fig:ucutResults}).

To avoid introducing more simulation parameters in the full 3D simulations than necessary, we account for the larger void volume created by the undercut by increasing the void diameter by \SI{60}{nm} (see Methods). 
Both for the merged and non-merged voids the experimental reflectance spectra agree well with the simulated spectra, indicating that the void modes are relatively robust against the change in void geometry, induced by the undercut, as long as the overall volume increase is taken into account. 

\subsection*{Near-field evidence of the void array modes}
The presence of multiple modes in a narrow spectral range makes it difficult to unambiguously identify each mode from far-field reflectance measurements.
To confirm the presence of the modes supported by the voids, we performed near-field measurements on three void arrays of diameters $d=360$~nm, $d=390$~nm, and $d=425$~nm excited at a wavelength of $\lambda=633$~nm (see Methods). By scanning an AFM tip of a scattering-type scanning near-field microscopy (SNOM) above the arrays, the topography of the arrays and the optical near-field intensity at each position of the scan are recorded simultaneously (Fig.~\ref{fig_4}). Since the AFM tip follows the sample topography, it also moves inside the voids, and thus scatters light from these. 
Figure~\ref{fig_4} shows the resulting near-field intensity of the void arrays, where the grey contour lines represent the edges of the holes, as extracted from the topography. 

For the the array with the smallest diameter voids, a high near-field intensity is observed inside the holes, while the near-field intensity is low inside the bridges (Fig.~\ref{fig_4}a). Both the excitation wavelength and the near-field profile matches with the resonance wavelength and the electromagnetic field of the coupled void mode (see Fig.~\ref{fig_3}c and Fig.~\ref{fig_2}h, respectively), supporting the interpretation that these observations are due to the excitation of the coupled void mode. As the void diameter increases to $d=390$~nm, we observe near-field intensity inside the voids as well as in the bridge regions (Fig.~\ref{fig_4}b). This is a sign of the spectral overlap between the coupled void and bridge modes. The near-field intensity from the void region is lowest for the array with diameter $d=425$~nm, where the bridge mode is primarily excited, confirming the declining presence of the void modes as the diameter increases. The near-field intensity from both the void and the bridge regions are highlighted by the three-dimensional (3D) representation in Fig.~\ref{fig_4}d, showing the 3D topography correlated with the near-field intensity. When exciting the void array with a longer wavelength that is off-resonant with all of the modes of the void array, we observe no near-field response from inside the voids~(Section~S7 and Fig.~\ref{fig:SNOM}). These near-field measurements are consistent with both the simulations as well as the far-field reflectance measurements, and confirm that the void arrays support both the coupled void mode and bridge mode. 

\subsection*{Void resonances in a fully encapsulated system}
The partially open voids support void modes with electromagnetic fields that extend outside the voids, suggesting that these modes are sensitive to encapsulation. To demonstrate this, we leverage van der Waals heterostructure assembly~\cite{geim2013van, pizzocchero2016} to encapsulate the void arrays. Encapsulation effectively reduces the void resonator volume, which shifts the fundamental void mode to shorter wavelengths. This brings two immediate advantages. First, at shorter wavelengths, the losses in WSe$_2$ are significantly larger, providing better confinement and less coupling between voids. Second, this shifts the void mode away from the bridge mode as the encapsulation is expected to have a minor effect on the bridge mode, and if any effect is observed, it would likely result in a red-shift.

We use two different layered materials, hBN and WSe$_2$, to encapsulate the voids. Details about the WSe$_2$ lid encapsulated devices are presented in the Supplementary Information~(Section~S9, Fig.~\ref{fig:WSe2Lid_SI}, and Fig.~\ref{fig:WSe2 cap}).
Although the refractive index of hBN is lower than that of WSe$_2$, hBN is transparent in the visible spectral range, allowing for unobstructed in-coupling of light into the encapsulated void. Simulations show that a hBN layer thickness of approximately 60$\,$nm is required to  confine the void modes~(see section S8 and Fig.\ref{fig:BNLid_SI}). The encapsulating lids are thin enough not to accommodate any optical modes themselves. Figure~\ref{fig_5}a shows a BF image of a hBN flake transferred onto voids of different dimensions, where significant differences in colour and contrast are observed between the regions with and without a hBN lid. Figure~\ref{fig_5}b shows the reflectance from the voids with (solid lines) and without hBN encapsulation (dashed lines). The collected spectra are from voids fabricated on two different flakes, where the top two spectra (marked by red stars) are the voids that are also shown in Fig.~\ref{fig_3}, with a 58$\,$nm thick hBN lid. The bottom three spectra (marked by blue stars) are collected from voids fabricated on a 1055$\,$nm thick WSe$_2$ flake and with a 53$\,$nm thick hBN lid.
The corresponding simulated reflectance of the hBN encapsulated voids is seen in Fig.~\ref{fig_5}c. For the smallest void array, the differences in the reflectance spectra obtained with and without the hBN lid are minimal. This is likely due to the limited capability of such small voids to effectively confine light. For larger voids, the reflectance dip is clearly visible and blueshifts when the hBN lid is added. As an example, the void array with diameter $d=370$~nm displays a wavelength shift of the reflectance dip of more than 150$\,$nm. This is a result of the reduced resonator volume in the encapsulated voids. Furthermore, when the void diameter increases, the reflectance dip associated with the fundamental void mode redshifts, as outlined by the thick red line in Fig.~\ref{fig_5}c. This behaviour is expected for void resonators with a growing volume. The simulated field distributions show that the void modes are present in both the open and encapsulated voids (Fig.~\ref{fig_5}d-e). Both field distributions are from the voids with $d = \SI{390}{nm}$ obtained at the reflectance dips, outlined by the black star and triangle in Fig. \ref{fig_5}b.
In the open void, a part of the electromagnetic field is localised inside the WSe$_2$ bridge (Fig.~\ref{fig_5}d). As described in Fig.~\ref{fig_2} and Fig.~\ref{fig_4}, this is likely a result of coupling between neighbouring voids. At the void resonance of the encapsulated voids (Fig.~\ref{fig_5}e) the magnitude of the field in the bridge region is decreased. This can be explained by the resonance being shifted towards shorter wavelengths, where the increased extinction coefficient of WSe$_2$ limits the coupling between neighbouring voids. 

\section*{Discussion}
We have demonstrated the formation of optical void modes confined in partially open and fully encapsulated void arrays using a lossy dielectric WSe$_2$, challenging the traditional notion that optical resonators require lossless dielectric materials. In the partially open system, we observe contributions from uncoupled and coupled void resonances as well as resonances in the bridge regions in reflectance measurements. These results are supported by near-field microscopy that directly visualise the void modes and the fields inside the WSe$_2$ bridge, in agreement with the finite-element simulations. Covering the voids with a lid reduces the effective resonator volume and shifts the void resonances to shorter wavelengths, where they do not overlap with other inherent modes of the system. The van der Waals heterostructure assembly makes it possible to use different layered materials to encapsulate the voids, as exemplified here using hBN and WSe$_2$, and thereby tailor the optical properties of the void system. As encapsulation shifts the void resonances to shorter wavelengths, this lessens the need for smaller voids to achieve similar effects, which may ease the initial fabrication process.
Our results highlight the versatility of void resonators in overcoming material constraints and emphasize the continued importance of exploring new material configurations. The ability to tailor optical modes by adjusting void dimensions along with the introduction of encapsulation layers open avenues for innovations in biomolecular sensing, metasurfaces, and other photonic devices.

\section*{Materials and Methods}

\textbf{Sample fabrication}
WSe$_2$ from HQ Graphene is mechanically exfoliated onto $\SI{90}{nm}$ SiO$_2$ on Si using thermal release tape to obtain thick flakes. 
The void arrays are defined by electron-beam lithography (EBL) in a \SI{30}{keV} Raith eLINE Plus system using a dose of \SI{230}{\micro C/cm^2} and an aperture of \SI{30}{\micro m}. A solution of \SI{4}{wt\%} PMMA 996 K in anisole is used as a resist (spun at \SI{2000}{rpm} for \SI{60}{s} and post-baked at \SI{170}{\celsius} for \SI{300}{s}). The resist is developed in IPA:H$_2$O (3:1) for \SI{60}{s} and rinsed in IPA for \SI{30}{s}. The voids are etched into WSe$_2$ using SF$_6$ reactive ion etching in a SPTS ICP system (pressure: \SI{10}{mTorr}, flowrate: \SI{40}{sccm}, coil/platen power: 0/30 W and time: 150 s). \newline
\textbf{Reflectance spectroscopy and bright field microscopy.} Reflectance spectra and bright field micrographs are obtained with a Nikon Eclipse LV100ND microscope with white-light illumination (Thorlabs OSL2).
Reflectance spectra are acquired with 20x objective (Nikon, NA$=$ 0.45) to ensure close to normal incident light and appropriate magnification (see Section~S2 and Fig.~\ref{fig:Reflection figs}).
An Andor Kymera 328i spectrograph is used to analyse the reflected light. The measured spectrum from the void arrays is normalised to the spectrum from a nearby unpatterned region of the WSe$_2$ flake.  
\newline
\textbf{Transfer of hBN and WSe$_2$ lids.} The exfoliated flakes are transferred to the void arrays using a standard dry-transfer technique with a polycarbonate (PC)/polydimethylsiloxane (PDMS) stamp and a transfer system from hq graphene. To eliminate the possibility of alterations in the optical response because of the polymer residues, we compared reflection measurements performed on partially open voids before and after the stacking process (see Section~S2 and Fig.~\ref{fig:Reflection figs}).
\newline
\textbf{Scanning electron microscopy (SEM).} SEM is performed in a Zeiss Supra VP 40 SEM under high vacuum and with a $\SI{2}{kV}$ acceleration voltage. Images acquired with the in-lens detector and zero tilt angle are used to determine the effective diameter of the hexagonal holes. Images acquired with the secondary electron detector and $\SI{40}{\degree}$ tilt angle are used to determine whether the voids have merged or not. \newline
\textbf{Atomic force microscopy (AFM).} The void depth is measured from AFM measurements acquired with a Dimension Icon-PT AFM from Bruker AXS in tapping mode.  \newline
\textbf{SNOM}. For the near-field measurements, we have used a scattering-scanning near-field optical microscope (s-SNOM, Attocube, neaSNOM) in reflection mode. The incident light from a tunable continuous-wave laser (Santec, TSL710) is focused on a platinum-coated silicon AFM probe (NanoWorld, Arrow-NC) with a nominal tip radius of 25~nm. The probe is used in intermittent contact mode at a frequency ${f_0}$=280 kHz. To minimise the perturbation from the tip, the laser is \textit{s}-polarized. We also use the cross-polarization technique to isolate the signal from the modes. Hence, a polariser (Thorlabs, LPNIR100-MP2) is placed in front of the photoreceiver (New Focus, 2053-FS) to select the \textit{p} polarisation of the scattered field. The recorded signal is demodulated with a lock-in amplifier at a frequency of ${4f_0}$. \newline
\textbf{Reflectance simulations.} Simulated reflectance spectra and electromagnetic field plots are obtained using the finite-element method in COMSOL Multiphysics (Wave Optics module).
The void arrays are modelled as circular cylindrical holes in a semi-infinite WSe$_2$ substrate (using a perfectly matched layer) and with a 500~nm period (realised with periodic boundary conditions). The voids and the region above WSe$_2$ is filled with air ($n$ = 1). 
The complex refractive index of WSe$_2$ is taken from ref. \cite{Munkhbat:2022}, with a modified value of $k$ in the out-of-plane $z$ direction ($k_{z} = 0.15$ instead of $k_{z} = 0$)~(see Section~S4 and Fig.~\ref{fig:SharpPeak_test_SI}). This is done to eliminate sharp peaks in the reflectance spectra that we do not observe experimentally. We believe the peaks arise due to sharp optical modes within the WSe$_2$, which may be hampered by defects induced during fabrication of the fabricated structures. 
The refractive index of hBN is taken from Ref.~\cite{Segura:2018}. The system is excited by a plane wave with normal incidence using a periodic port placed at a distance $\lambda/2$ above the WSe$_2$.
The reflectance from the void array $R_\textrm{void}$ is normalised to the reflectance from a plane air-WSe$_2$ interface $R_\textrm{flat}$ to obtain the reflectance of the void system, $R = R_\textrm{void}/R_\textrm{flat}$. All reflectance curves correspond to zeroth-order diffraction.
Field plots and reflectance maps from simulations as well as near-field intensity scans from SNOM are plotted with colour maps from Ref.  \cite{CrameriSOFTWARE:2018,CrameriPAPER:2020}.\newline
\textbf{Eigenmode simulations.}
Eigenmode simulations are performed with a similar simulation geometry using the ARPACK Eigenfrequency solver in COMSOL Multiphysics (Wave Optics module). Because the eigenmode analysis requires constant refractive indices, we set the refractive index components of WSe$_2$ to the average values in the wavelength range of interest (500-760~nm).
The in-plane components of $n$ are $n_{x} = n_{y} = \SI{4.6476}{}$, the out-of-plane component is $n_{z} = 2.9423$, while the in-plane components of $k$ are $k_{x} = k_{y} = 1.2242$. The out-of-plane component of $k$ is again set to $k_{z}=0.15$ to eliminate sharp optical modes within the WSe$_2$. 

\bibliography{scibib}

\bibliographystyle{Science}

\section*{Acknowledgments}
The authors also thank Matías Vázquez for his assistance in tilted SEM measurements.

\subsection*{Funding}
We acknowledge support by the Independent Research Funding Denmark (1032-00496B).

\subsection*{Author contributions}

\textbf{Avishek Sarbajna:} Sample preparation and device fabrication, far-field optical measurements and data analysis, theoretical framework, AFM measurements, writing original draft. \textbf{Dorte Rubæk Danielsen:} Sample preparation, Comsol simulations and data analysis, theoretical framework, AFM measurements, SEM measurements, writing original draft. \textbf{Laura Casses:} Near-field optical measurements and data analysis, writing original draft. \textbf{Nicolas Stenger:} Co-supervision, writing original draft. \textbf{Peter Bøggild:} Co-supervision, writing original draft. \textbf{Søren Raza:} Supervision, writing original draft, conceived idea, SEM data analysis.

\subsection*{Competing interests}
The authors declare that they do not have any known competing financial interests or personal relationships that might have influenced the findings presented in this paper.
\subsection*{Data and materials availability}

All the data and figures used in the paper, along with supplementary data, are available from the corresponding authors upon request.

\newpage
\section*{Main Figures}

\begin{figure}[H]
    \includegraphics[width=\textwidth]{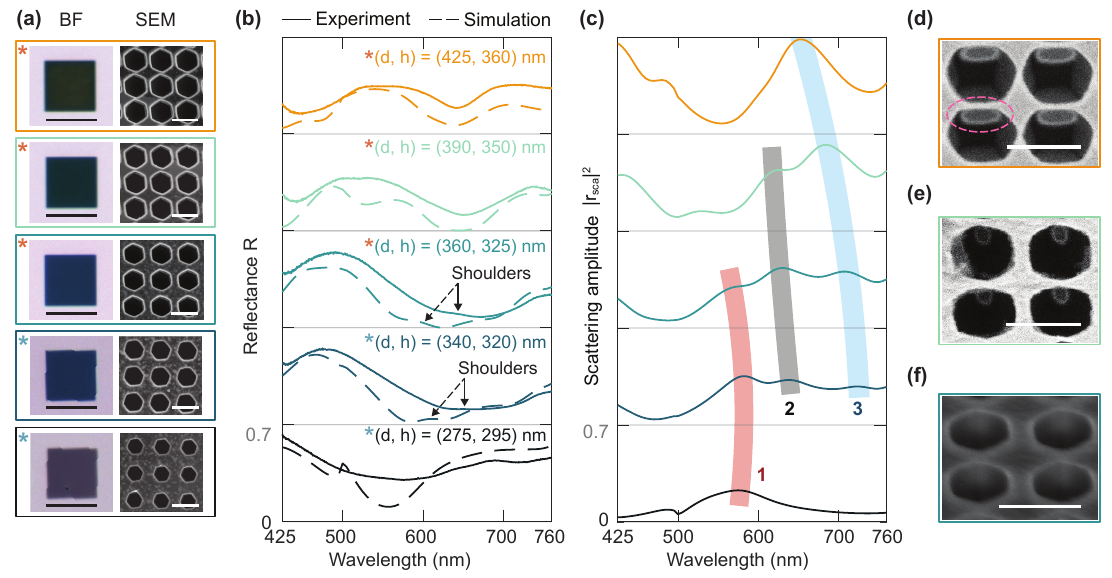}
    \caption{\textbf{Resonances in densely packed open WSe$_2$ voids} \textbf{(a)} BF (left) and SEM (right) images of void arrays etched into two different WSe$_2$ flakes with thicknesses of \SI{438}{nm}  (orange stars) and \SI{422}{nm} (blue stars), respectively. Scale bars in BF images: \SI{15}{\micro m}. Scale bars in SEM images: \SI{500}{nm}. \textbf{(b)} Experimental (solid) and simulated (dashed) reflectance spectra from void arrays depicted in (a) with diameter $d$ and depth $h$. In simulations, the void diameter is \SI{60}{nm} larger to accommodate the observed undercut. \textbf{(c)} Scattering amplitude $\abs{r_{\textnormal{sca}}}^2$ extracted from the simulated reflection coefficients. Thick lines outline the peaks corresponding to the modes: (1) void mode, (2) coupled void mode, and (3) bridge mode. \textbf{(d-f)} Tilted SEM images of the voids with $d = \SI{425}{nm}$, $d = \SI{390}{nm}$, and $d = \SI{360}{nm}$, respectively. The voids in (d) and (e) are merged as a result of the etched undercut creating a small hole in the sidewall between the voids (pink dashed line). The voids in (f) are not merged. Scale bars: \SI{500}{nm}. 
    }
    \label{fig_3}
\end{figure}

\begin{figure}[H]
    \includegraphics[width=\textwidth]{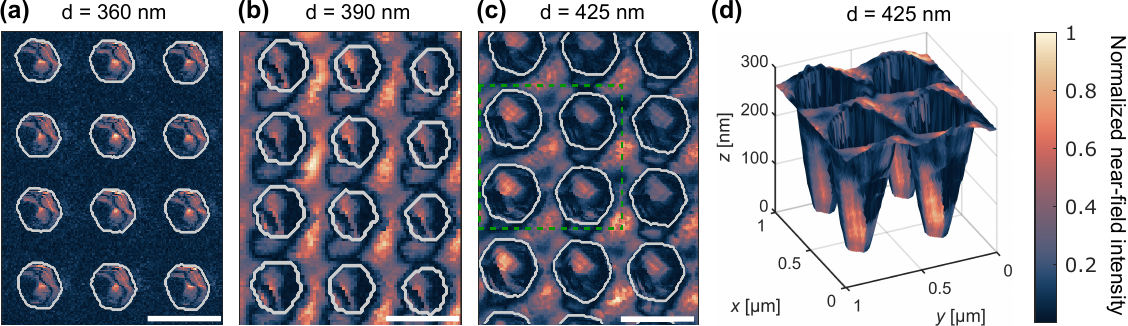}
    \caption{\textbf{Near-field measurements of void arrays}. \textbf{(a-c)} Near-field intensity at the wavelength $\lambda = 633$~nm for void arrays with diameters of (a) $d = 360$~nm, (b) $d = 390$~nm, and (c) $d = 425$~nm. The grey lines on the map show the position of the holes based on the topography map. \textbf{(d)} Three-dimensional representation of the subset of holes indicated by the green dotted lines in (c). Scale bars: 500 nm.}
    \label{fig_4}
\end{figure}

\begin{figure}[H]
    \includegraphics[scale = 0.95]{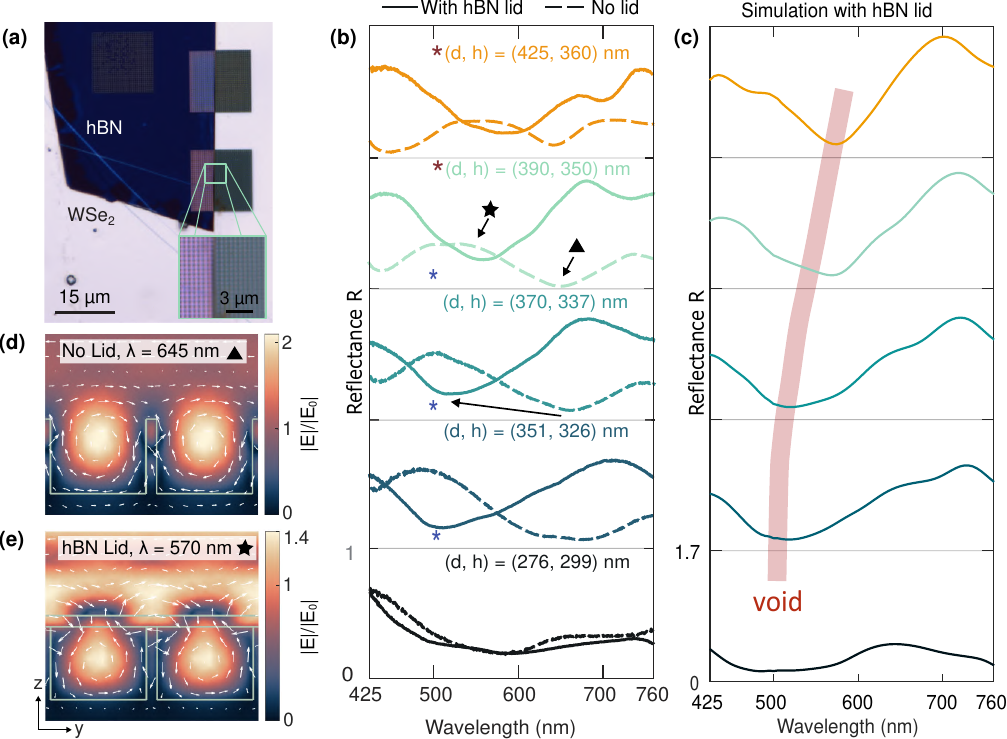}
    \caption{\textbf{Experimental realization of encapsulated void arrays.} \textbf{(a)} BF micrographs of void arrays in WSe$_2$ covered with a hBN lid. Inset: zoom-in of the partially covered void array. \textbf{(b)} Experimental reflectance spectra from the voids shown in (a) with (solid) and without (dashed) an hBN lid. Spectra from different flakes are identified by red and blue stars. \textbf{(c)} Corresponding simulated reflectance spectra for the hBN encapsulated voids. The red line indicates the void mode resonance wavelength. \textbf{(d-e)} Simulated electric (colour map) and magnetic (white arrows) field distributions in the voids with $d = \SI{390}{nm}$ without (d) and with a hBN lid (e), obtained at the reflectance dips outlined by the black triangle ($\lambda = \SI{645}{nm}$) and black star ($\lambda = \SI{570}{nm}$) in (b).}
    \label{fig_5}
\end{figure}
\newpage

\begin{center}
\section*{Supplementary Material for}

\section*{Encapsulated void resonators in lossy dielectric
van der Waals heterostructures} 
\author
{Avishek Sarbajna \textit{et al.}}
\normalsize{$^\ast$Corresponding author: S{\o}ren Raza, E-mail: sraz@dtu.dk}
\end{center}

\section*{Section S1: Simulations with different flake thicknesses}
In the main text, we measure the reflectance from void arrays fabricated on exfoliated WSe$_2$ flakes with different initial thicknesses. The varying thicknesses of the flakes leaves residual WSe$_2$ below the voids with thicknesses $t$ in the range from 55~nm to 125~nm. To confirm that the reflectance from the voids does not depend on the residual thickness $t$, we performed additional 3D COMSOL simulations of voids with various WSe$_2$ residual thicknesses. In the simulation geometry we also include the substrate of \SI{90}{nm} SiO$_2$ and semi-infinite Si (modelled using a PML layer) as illustrated in Fig.~\ref{fig:FlakeThicknessSweep_and_SiO2SiSubsComp_SI}(a). The simulated reflectance spectra from voids with increasing residual thicknesses of WSe$_2$ is shown in Fig. \ref{fig:FlakeThicknessSweep_and_SiO2SiSubsComp_SI}(b). We observe that there is no considerable spectral change in the experimentally relevant range of $t=55$~nm to $t=125$~nm.  Fig.~\ref{fig:FlakeThicknessSweep_and_SiO2SiSubsComp_SI}(c) shows a comparison between the reflectance from voids of the same size with $t=90$~nm (yellow line) and with an infinite WSe$_2$ substrate (black line). The two spectra are almost identical. Therefore we neglect the SiO$_2$-on-Si substrate in all other simulations and model the WSe$_2$ layer as infinitely thick using a PML layer. 

\section*{Section S2: Reflectance Measurements}

To study the far-field response of the Mie voids we have performed reflectance spectroscopy (see Methods). In this section we will cover several details related to reflectance spectroscopy results shown in the main text.  

\subsection*{Comparison of different objectives}
To achieve quasi-normal incident and collected light, 
it would be ideal to use objectives with low numerical apertures. However, to ensure that the collected light is only from the region of interest, certain magnification is required. In Fig. \ref{fig:Reflection figs}(a), three reflectance spectra from the same void array (with diameter $d = \SI{360}{nm}$ and depth $h = \SI{325}{nm}$) obtained with three different objectives are shown. The spectra taken with 10x (NA = 0.30) and 20x (NA = 0.45) objectives (yellow and red line) are almost identical, whereas the spectrum taken with the 50x (NA = 0.80) objective starts to deviate. Therefore we used 20x objective throughout the measurements.   

\subsection*{Impact of the stacking process}
To transfer lids onto the void arrays, we use a stacking technique based on polycarbonate (see Methods). After stacking left-over polycarbonate is dissolved in chloroform and the sample is cleaned in isopropyl alcohol and acetone. To confirm that the stacking process itself, where the sample is in contact with polymers, does not affect the optical response, we performed reflection measurements on a void array, which was not covered by the lid before and after the stacking. The reflectance measured before and after stacking completely match, as seen in Fig. \ref{fig:Reflection figs}(b). This shows that the optical response is not impacted by the stacking process itself.  

\subsection*{Impact of flake thickness}
Different flake thicknesses lead to different residual thicknesses of WSe$_2$ underneath the voids. To check whether different thicknesses impact the optical response of the voids, we compared the reflectance of void arrays fabricated on \SI{422}{nm}, \SI{438}{nm}, and \SI{1055}{nm} thick flakes. The void diameters were kept in the same range (\SI{377}{nm}, \SI{360}{nm}, and \SI{370}{nm}, respectively) to minimise the effect of change in dimension. As seen in Fig. \ref{fig:Reflection figs}(c), the reflectance is larger overall for the 1055~nm thick flake, but the spectral shape is the same in all three cases. The slight spectral deviations between the flakes of different thicknesses are attributed to the void dimensions (diameter and depth) not being identical.

\section*{Section S3: Effect of reactive ion etching on void dimension}

For sufficiently small features, it is well-known that the etch rate in reactive ion etching can depend on the feature sizes of the etched pattern \cite{Danielsen:2021,lee:1991a,gottscho:1992a}.
Consequently, we observe that etched voids of different diameters have different depths. As seen in Fig. \ref{fig:Diam_vs_depth} the depth increases linearly with the diameter for the voids fabricated in flakes of different thicknesses.

\section*{Section S4: Origin of sharp peaks in 3D simulations}

In the simulations we use the anisotropic refractive index of WSe\textsubscript{2} from ref. \cite{Munkhbat:2022}. The imaginary part of the out-of-plane refractive index $k_{\mathrm{z}}$ is zero for all wavelengths in ref. \cite{Munkhbat:2022}. However, when 3D simulations of the voids were performed with $k_{\mathrm{z}} = 0$, sharp features appear in the reflectance spectra (see Fig. \ref{fig:SharpPeak_test_SI}(a)). These features are not visible in the experimental results. To understand the origin of these sharp peaks we performed simulations with various void geometries (see Fig. \ref{fig:SharpPeak_test_SI}(a)), keeping the total void volume as unaltered as possible. The sharp peaks persist for voids with rough side walls (pink line) and hexagonal holes (green line). The spectra from these three void geometries are almost identical, which also proves the voids are robust against changes in the geometry. 
For the hexagonal skewed void (blue line) and the void with an elliptical undercut (yellow line) the sharp peaks are modified, but are still present. 
Fig. \ref{fig:SharpPeak_test_SI}(b,c) illustrates the simulation geometries for the rough circular void and void with an undercut, respectively. 
In contrast to the changes in the void geometry, increasing the value of $k_{\mathrm{z}}$ dramatically reduces the sharp reflectance peaks, as seen from Fig. \ref{fig:SharpPeak_test_SI}(d). The field distributions in Fig. \ref{fig:SharpPeak_test_SI}(e,h) obtained at a sharp peak ($\lambda =  \SI{610}{nm}$) provide an explanation for this. Fig. \ref{fig:SharpPeak_test_SI}(e,f) show the $(x,y)$ cross sections at $z = h/2$ of the electric field (colour map) and magnetic field (arrows), while Fig. \ref{fig:SharpPeak_test_SI}(g) and (h) show $(x,z)$ cross sections at $y = 0$ of the magnetic field (colour map) and electric field (arrows).
When $k_{\mathrm{z}} = 0$ the field distribustions in Fig. \ref{fig:SharpPeak_test_SI}(e,g) suggest that a mode is present within the WSe\textsubscript{2} bridge with strong electric and magnetic field enhancement in the bridge region and the electric field pointing in the $z$-direction (i.e., out-of-plane direction).
Because of the direction of the electric field, this mode gets suppressed when $k_{\mathrm{z}}$ increases slightly. This is seen in Fig. \ref{fig:SharpPeak_test_SI}(f,h) where $k_{\mathrm{z}} = 0.15$ and the electric field is no longer predominantly in the $z$-direction.
In the experiments, we suspect that the etching process and the observed undercut could induce defects that effectively lead to an increase in $k_{\mathrm{z}}$ and thereby masks these sharp resonances in the experimental measurements.

\section*{Section S5: Topography of the void arrays}
To determine the effective void diameter of the hexagonal voids, we process the SEM images in MATLAB as illustrated in Fig.~\ref{fig:Diam_SEM}. The contrast of the original image (Fig. \ref{fig:Diam_SEM}(a)) is first enhanced (Fig. \ref{fig:Diam_SEM}(b)). By setting a threshold grey-scale value, the image is converted into a binary black-and-white image (Fig. \ref{fig:Diam_SEM}(c)). For convenience, the image is inverted so the void regions are white (Fig. \ref{fig:Diam_SEM}(d)). Regions with voids are detected (Fig. \ref{fig:Diam_SEM}(e)) and their areas $A$ are calculated by determining the number of pixels within each void and multiplying by the pixel area. The equivalent diameters are then calculated as $d = \sqrt{4A/\pi}$, and the mean diameter of all voids within the SEM image is reported.   

Figure~\ref{fig:topography AFM}(a) shows an AFM measurement of a single void. The bottom of the void has minimum roughness and can be considered flat. 
The void depth is measured from line scans across the void as illustrated in Fig.~\ref{fig:topography AFM}(b).

\section*{Section S6: Modelling void undercut in 2D simulations}

The etched WSe\textsubscript{2} voids have undercuts, as seen from SEM images (see Main Fig.3(d-f)). Because it is difficult to know the exact undercut geometry, it is taken into account in simulations by increasing the void diameters by 60 nm compared to the measured values. Furthermore, the largest voids are merged as a result of the undercut. However, the far-field (Main Fig. 3) and near-field response (Main Fig. 4) suggest that the merged voids follow the same trends as the non-merged voids. 

To understand the impact of the undercut on the optical response of the voids, we performed 2D COMSOL simulations of voids with and without undercuts.
The simulation geometry is illustrated in Fig. \ref{fig:ucut_sim_geometry}. 
The undercut shape is created by superimposing an ellipse onto the void.
The ellipse has semi-major axis $a$, which increases for increasing undercut widths, and semi-minor axis $b$ which is held constant at 133 nm.
We define a parameter $\delta$ that quantifies the degree of merging as $\delta = \Lambda/2 - a$, where $\Lambda$ is the array period. For non-merged voids, $2\delta$ is the minimum bridge width and $\delta>0$. For merged voids, $\delta$ defines the part of the ellipse that extends beyond the simulation domain and $\delta<0$.
The voids with undercuts are compared to those without undercuts with the same area by keeping the depth constant and changing the width so $d = A_{\mathrm{undercut}}/h$.

The reflectance spectra from voids with undercuts and varying $\delta$-values are shown in Fig. \ref{fig:ucutResults}(a), while the spectra from the corresponding voids without undercuts are shown in Fig. \ref{fig:ucutResults}(b). The spectra appear to be similar from $\delta = \SI{95}{nm}$ to $\delta \approx \SI{-20}{nm}$. 
To compare the reflectance directly, Fig. \ref{fig:ucutResults}(c) shows individual spectra from voids with undercuts (solid lines) and voids without undercuts (dashed lines).
The field distributions in Fig. \ref{fig:ucutResults}(d)-(f) represent the lowest-wavelength void mode in voids with varying undercuts. 
When $\delta = \SI{50}{nm}$ the reflectance spectra in Fig \ref{fig:ucutResults}(c) match well. The field distributions in voids with and without undercuts are also very similar (Fig. \ref{fig:ucutResults}(d)). When $\delta = 0$, the reflectance dips are still at the same positions, but the intensity varies slightly. The field distributions are still very similar (see Fig. \ref{fig:ucutResults}(e)).
Finally, when $\delta = \SI{-35}{nm}$ the reflectance spectra start to differ more. Even though the reflectance dip positions are unchanged, the dip associated with the lowest-wavelength void mode is almost gone, while the higher-wavelength mode is much more pronounced for the merged voids compared to the non-merged voids. As seen in Fig. \ref{fig:ucutResults}(f), the mode profile in the merged voids starts to differ slightly and the field intensity is lower, compared to the voids without undercuts. 

The agreement between voids with an undercut and non-undercut voids with a larger diameter demonstrates that changes to the void mode induced by the undercut can be modelled by an increase in void diameter.

\section*{Section S7: Additional near-field measurements on and off resonance}

Fig. \ref{fig:SNOM} shows a comparison between near-field measurements of the same void array obtained with an excitation wavelength of 633~nm (near the bridge mode resonance) and at an excitation wavelength of 730~nm (off resonance). The voids are 425~nm in diameter (see Fig. \ref{fig_3} (yellow lines) and Fig. \ref{fig_4} (c-d)).

At the bridge resonance (Fig.~\ref{fig:SNOM}(a-b)), the near-field intensity is high inside both the voids and the bridge regions. 
When the excitation wavelength is off resonance (Fig.~\ref{fig:SNOM}(cd)) there is no high intensity features inside the voids, as highlighted by the 3D representation in Fig. \ref{fig:SNOM}(d). This proves that the high intensity features in (a) and (b) are not related to topography artefacts, but is a signature of the void array supporting resonant modes.

\section*{Section S8: Optimal lid thickness}

To assess which lid thicknesses are suitable for void mode confinement, we perform 2D simulations of voids encapsulated by lids of varying thicknesses.
Fig. \ref{fig:BNLid_SI}(a) shows a schematic illustration of the 2D simulation geometry of a void array encapsulated by an hBN lid. There are periodic boundary conditions on the left and right side and a perfectly matched layer at the bottom to model bulk WSe\textsubscript{2}. As the hBN lid thickness increases, the reflectance increases overall, while a reflectance dip appears around \SI{620}{nm} (see Fig. \ref{fig:BNLid_SI}(b)). The reflectance is normalized to the reflectance of bulk WSe\textsubscript{2}. As a proxy for the electromagnetic field enhancement in the void, we define the energy enhancement $W_{\mathrm{void}}/W_{\mathrm{0}}$ as the total electromagnetic energy density $u$ stored within the void and the energy density of the incident electromagnetic field $u_0$ integrated over the same volume: $W_{\mathrm{void}}/W_{\mathrm{0}} = \int_{void}u(r)\mathrm{d}r/\int_{void} u_0(r)\mathrm{d}r$. Fig. \ref{fig:BNLid_SI}(c) shows the energy enhancement for increasing hBN lid thicknesses. When there is no hBN lid, the energy enhancement peaks around \SI{680}{nm}. As the lid thickness increases the peak at \SI{680}{nm}, gradually decreases, while a new peak forms at shorter wavelengths, corresponding to the smaller mode volume of the encapsulated void. For \SI{60}{nm} thick hBN lids the energy enhancement exhibits a large peak around \SI{620}{nm}. For thicker lids, the peak starts to redshift and decrease. This suggests that hBN lids with thicknesses around \SI{60}{nm} are most suitable for void encapsulation.  

\section*{Section S9: Encapsulation with WSe$_2$ lid}
The simulation results for voids encapsulated with WSe\textsubscript{2} are shown in Fig.~\ref{fig:WSe2Lid_SI}. The 2D simulation geometry (Fig.~\ref{fig:WSe2Lid_SI}(a)) is similar to the one with hBN lids. Because the optical loss in WSe\textsubscript{2} is large in the visible range (due to the large imaginary part of the refractive index) the lids need to be thinner than the hBN lids. The reflectance and energy enhancement spectra (Fig.~\ref{fig:WSe2Lid_SI}(b-c)) show that the void resonance shifts to shorter wavelengths when the WSe\textsubscript{2} lid is added. When the lid is 10-15 nm thick the energy enhancement has a single peak around \SI{620}{nm}. The energy enhancement peak decreases for thicker lids and the overall reflectance increases. This suggests that the WSe\textsubscript{2} lid reflects a large portion of the incident light so it does not enter the void. 
Therefore we find that WSe\textsubscript{2} lid thicknesses around 10-\SI{15}{nm} are most suitable for void encapsulation.

We experimentally verified these features by encapsulating the voids with a thin WSe$_2$ lid (Fig.~\ref{fig:WSe2 cap}). The voids have dimensions similar to those  discussed in the main text, with array 1 featuring the smallest and array 3 the largest voids. The BF images (see Fig.~\ref{fig:WSe2 cap}(a)) also show a colour change upon encapsulation; however, the distinction between encapsulated voids of different sizes is less pronounced. The WSe$_2$ lid contains two regions of 9~nm and 16~nm thickness, respectively (see inset, Fig. \ref{fig:WSe2 cap}(a)). The void area covered by a 9$\,$nm lid imparts a light violet colour, while the \SI{16}{nm} lid region is a brighter pink. Consequently, the colour of the encapsulated array resembles the colour of the lid itself (see Fig.\ref{fig:WSe2 cap}(a)). We attribute this to the high refractive index and optical loss in the WSe$_2$ lid, which gives rise to absorption of light and increased light reflection from the lid surface. Nevertheless, reflectance measurements of the voids reveal a blueshift of the reflectance dips after encapsulation (Fig. \ref{fig:WSe2 cap}(b)). Furthermore, when comparing the reflectance from array 3, where one region is covered by a 16$\,$nm thick lid and the other by a 9$\,$nm thick lid, the thicker lid yields a less pronounced reflectance dip (see Fig.~\ref{fig:WSe2 cap}(c)). In addition, with the thicker lid there is an overall increase in reflectance compared to the thinner lid, which we attribute to an increased reflection from the WSe$_2$ lid surface.

\newpage


\renewcommand{\thefigure}{S\arabic{figure}}
\setcounter{figure}{0} 

\begin{figure}[H]
    \centering
    \includegraphics[width=\textwidth]{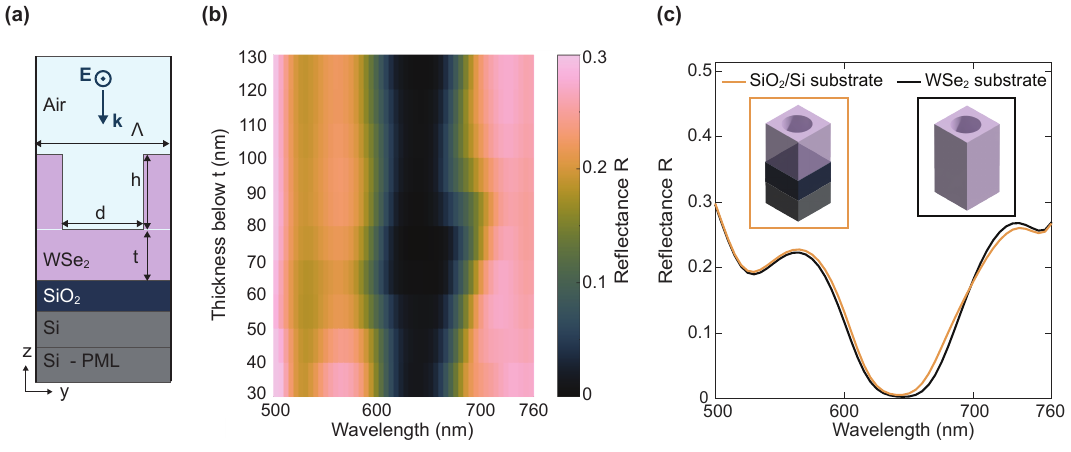}
    \caption{\textbf{Reflectance with varying flake thickness below the void}.
    \textbf{(a)} Schematic illustration of the simulation geometry, shown in 2D for clarity. The actual simulation geometry is three-dimensional. The WSe$_2$ is on a substrate of SiO$_2$ (90 nm thick) and Si (modelled as infinite). \textbf{(b)}
    Simulated reflectance spectra for a void array with $d = \SI{450}{nm}$, $h = \SI{350}{nm}$, and $\Lambda = \SI{500}{nm}$ in WSe\textsubscript{2} of varying thickness. \textbf{(c)} Simulated reflectance spectra from voids in semi-infinite WSe$_2$ (black line) and voids with a residual WSe$_2$ thickness 90 nm WSe$_2$ on a 90 nm SiO$_2$/Si substrate (yellow line). The voids have same dimensions as in (b).}
    \label{fig:FlakeThicknessSweep_and_SiO2SiSubsComp_SI}
\end{figure}

\begin{figure}[H]
    \centering
    \includegraphics[scale = 0.68]{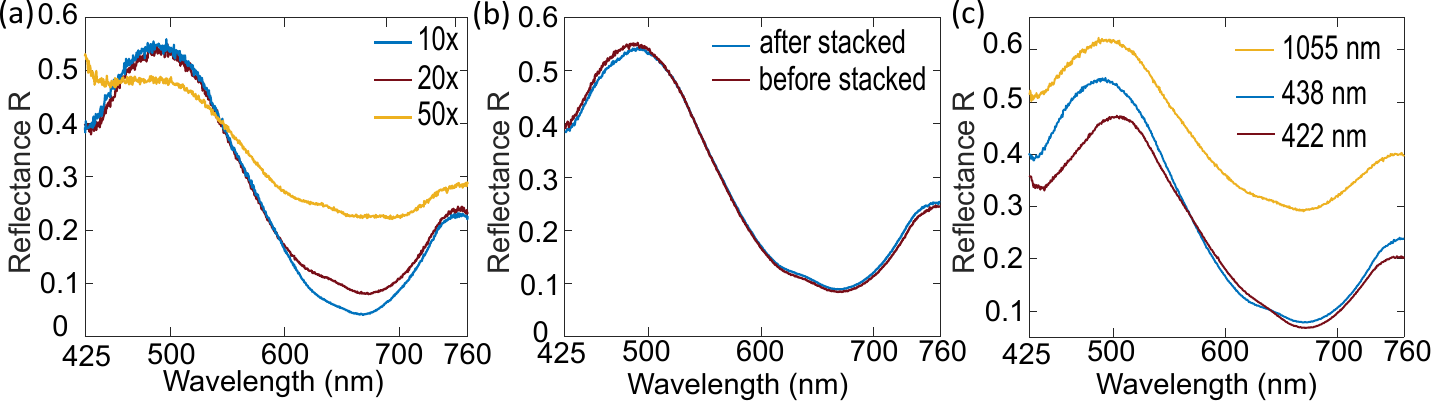}
    \caption{\textbf{Reflectance from void arrays.} \textbf{(a)} Reflectance measurements on void array with diameter $d=360$~nm and depth $h=325$~nm using objective lenses with different numerical apertures. \textbf{(b)} Reflectance measurements of uncovered void array before and after the stacking process ($d = \SI{360}{nm}$ and $h = \SI{325}{nm}$). \textbf{(c)} Reflectance measurements of void arrays fabricated on flakes with different thicknesses. The void diameters are 377~nm, 360~nm, and 370~nm fabricated on 422~nm, 438~nm, and 1055~nm thick flakes, respectively.}
    \label{fig:Reflection figs}
\end{figure}

      \begin{figure}[H]
    \centering
    \includegraphics[scale = 1.5]{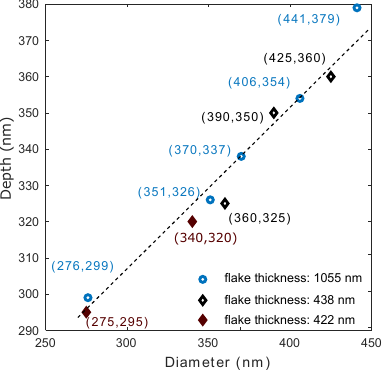}
     \caption{\textbf{Measured void diameter vs. depth.} The etched depth $h$ increases with increasing diameter $d$. The values in the bracket show the $(d,h)$ values of the voids. The black dotted line indicates the linear relation between the diameter and the depth.}
    \label{fig:Diam_vs_depth}
\end{figure}

\begin{figure}[H]
    \centering
    \includegraphics[scale = 0.95]{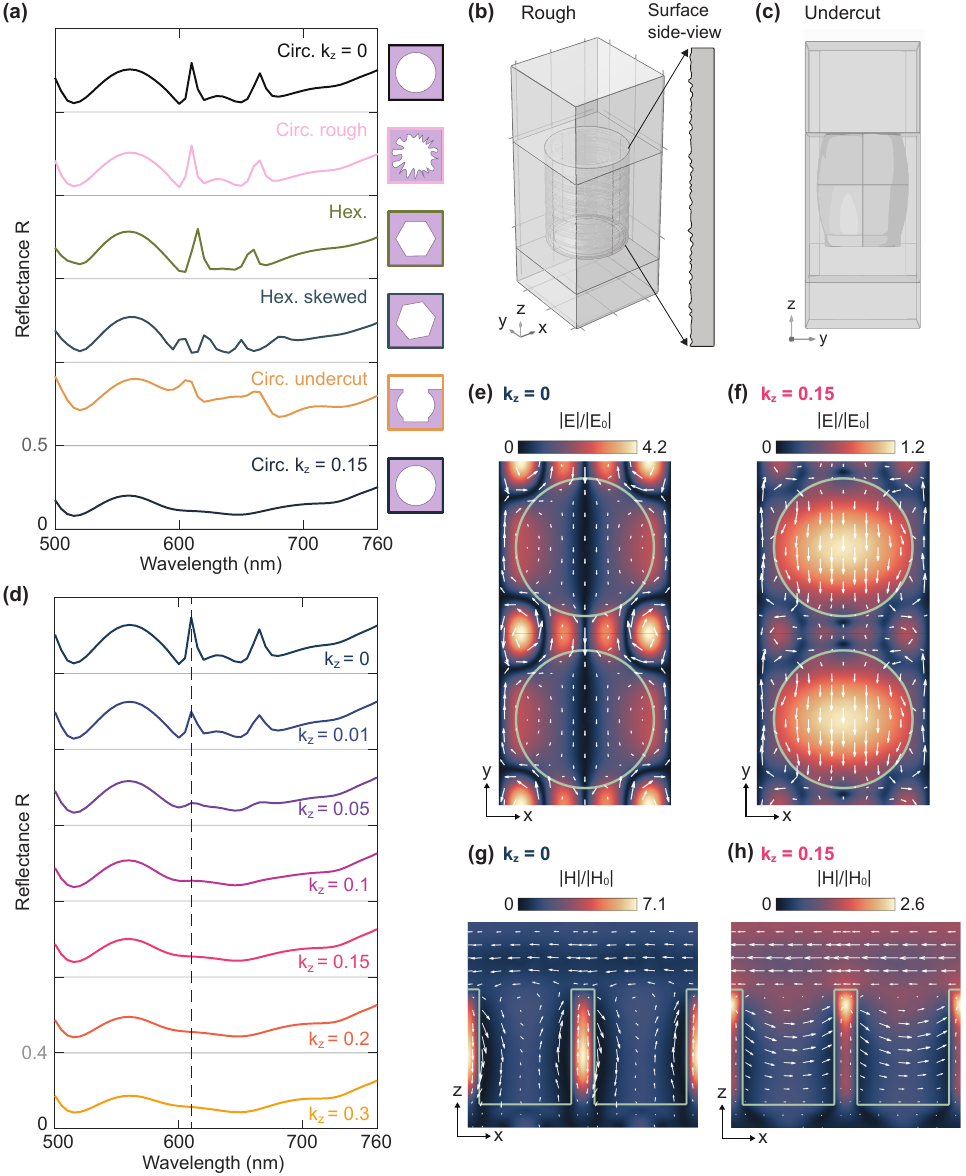}
    \caption{\textbf{Origin of sharp peaks. (a)} Reflectance spectra from various 3D void geometries. Circular holes have a 400 nm diameter and a depth of 500 nm. Hexagonal holes have the same depth and same cross-sectional area as the circular holes. The undercut geometry is made by superimposing an ellipsoid on the cylindrical hole. The ellipsoid has semiaxes $a = b = \SI{205}{nm}$ and $c = \SI{400}{nm}$. \textbf{(b-c)} Schematic illustration of the simulation geometry of the circular hole with a rough surface (b) and the undercut geometry (c). \textbf{(d)} Reflectance spectra from circular holes with increasing $k_z$-values. \textbf{(e-h)} Electric and magnetic field distributions at $\lambda = \SI{610}{nm}$ corresponding to the sharp peak (see dashed line in (d)) when $k_z = \SI{0}{}$ and $k_z = \SI{0.15}{}$.}
    \label{fig:SharpPeak_test_SI}
\end{figure}

   \begin{figure}[H]
         \centering
         \includegraphics[width = \textwidth]{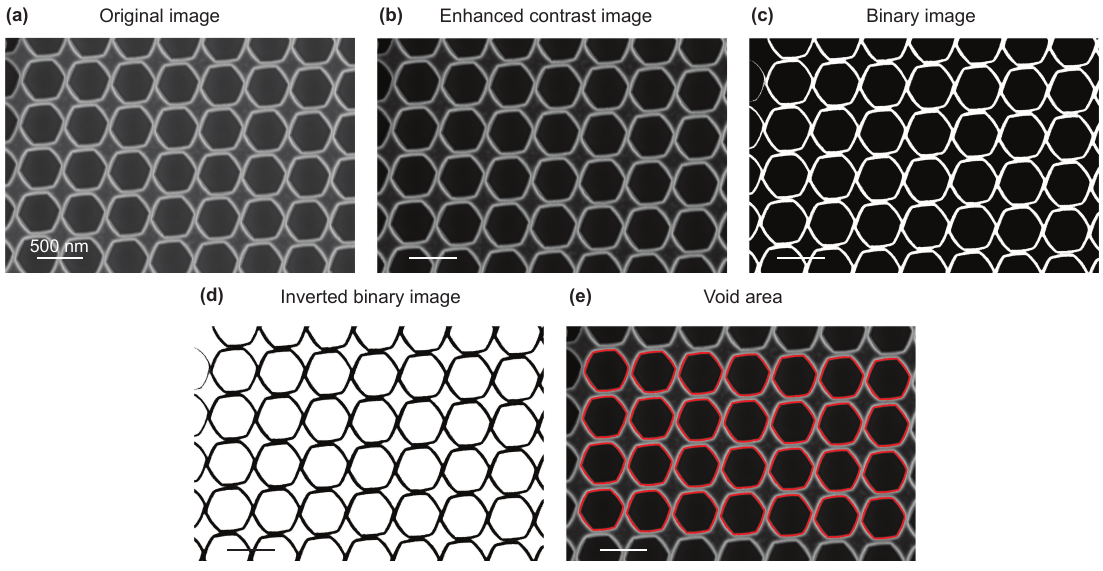}
         \caption{\textbf{Measuring the equivalent void diameter from SEM.} \textbf{(a)} Original SEM image. \textbf{(b)} SEM image with enhanced contrast. \textbf{(c)} Binary image. \textbf{(d)} Inverted binary image. \textbf{(e)} Original SEM image with the void edges are outlined with red lines. The void edges are detected from the inverted binary image. Scale bars: \SI{500}{nm}.}
         \label{fig:Diam_SEM}
     \end{figure}

\begin{figure}[H]
         \centering
        \includegraphics[width = 0.9\textwidth]{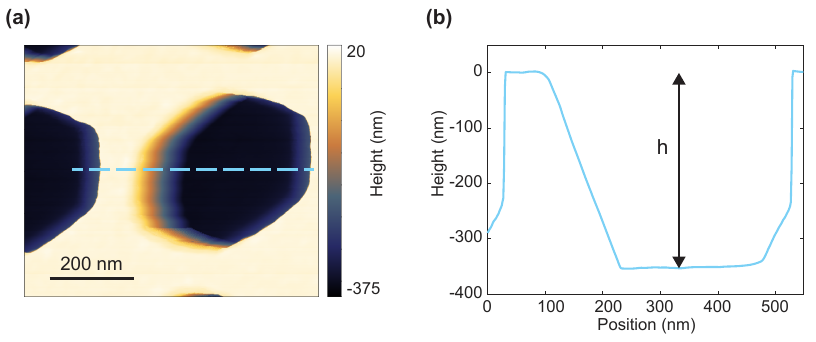}
         \caption{\textbf{Measuring the void depth from AFM}. \textbf{(a)} AFM scan of a single void. \textbf{(b)} Line scan across the void indicated by the red dashed line in (a). The void depth $h$ is shown.}
         \label{fig:topography AFM}
         \end{figure}

\begin{figure}[H]
    \centering
    \includegraphics{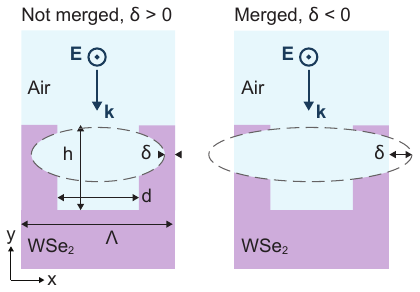}
    \caption{\textbf{Simulation geometry for voids with an elliptic undercut.} Simulations are performed with normal incidence light polarized in the $z$-direction, and void depth $h = \SI{400}{nm}$, diameter $d = \SI{350}{nm}$, and period $\Lambda=\SI{500}{nm}$. For non-merged voids, $\delta>0$ and $2\delta$ denotes the minimum bridge width at the undercut. For merged voids $\delta<0$.}
    \label{fig:ucut_sim_geometry}
\end{figure}

\begin{figure}[H]
    \centering
    \includegraphics[scale=0.93]{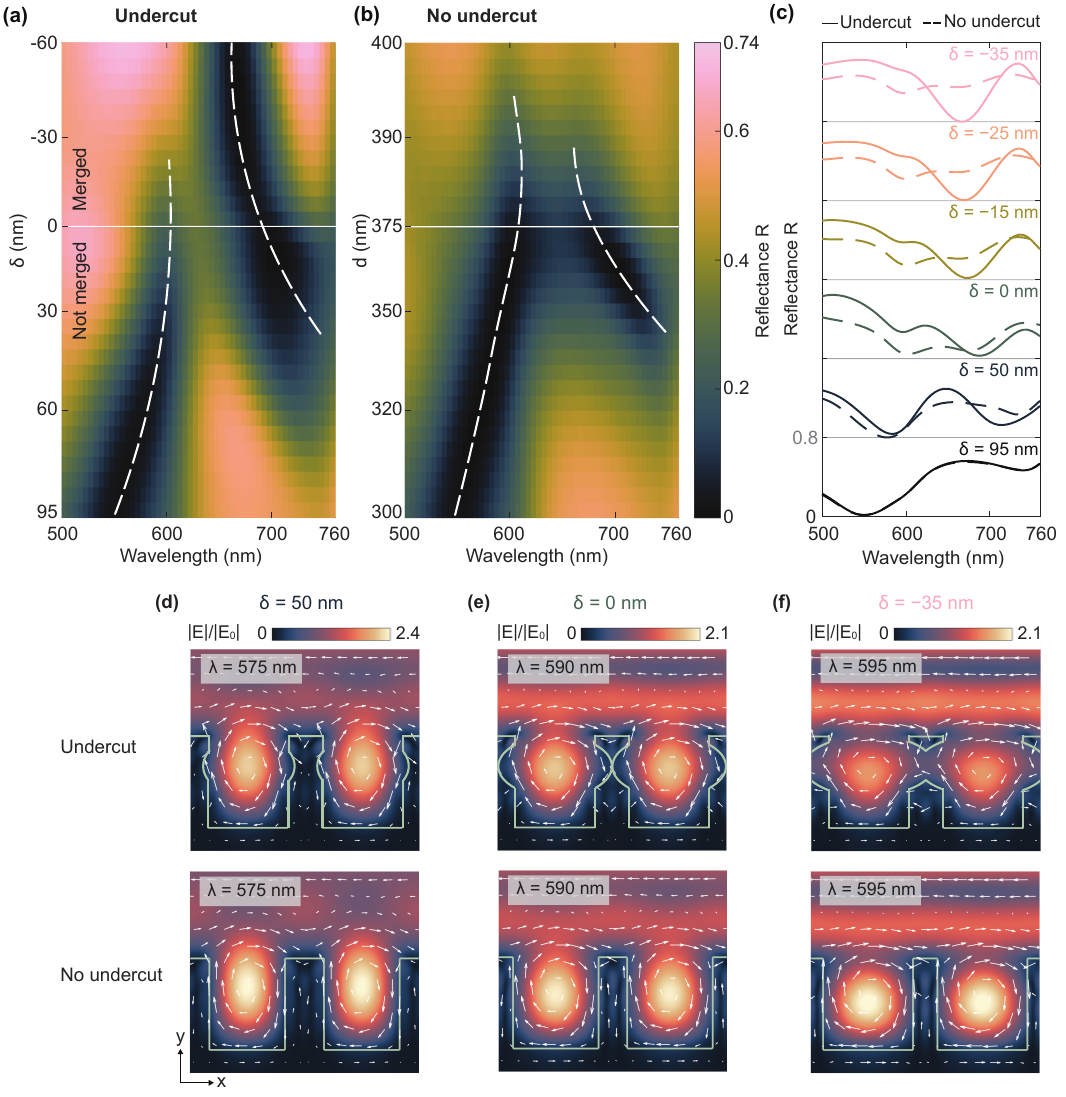}
    \caption{\textbf{Impact of undercut on void resonances.} \textbf{(a)} Reflectance from 2D voids with increasing undercuts. \textbf{(b)} Reflectance from 2D voids without undercut but with the same area as in (a). \textbf{(c)} Individual spectra from several voids with varying $\delta$. \textbf{(d-f)} Electric (colour maps) and magnetic (white arrows) field distributions at different wavelengths along the lowest-wavelength reflectance dip for voids with and without undercuts.}
    \label{fig:ucutResults}
\end{figure}

\begin{figure}[H]
    \centering
    \includegraphics[width=\textwidth]{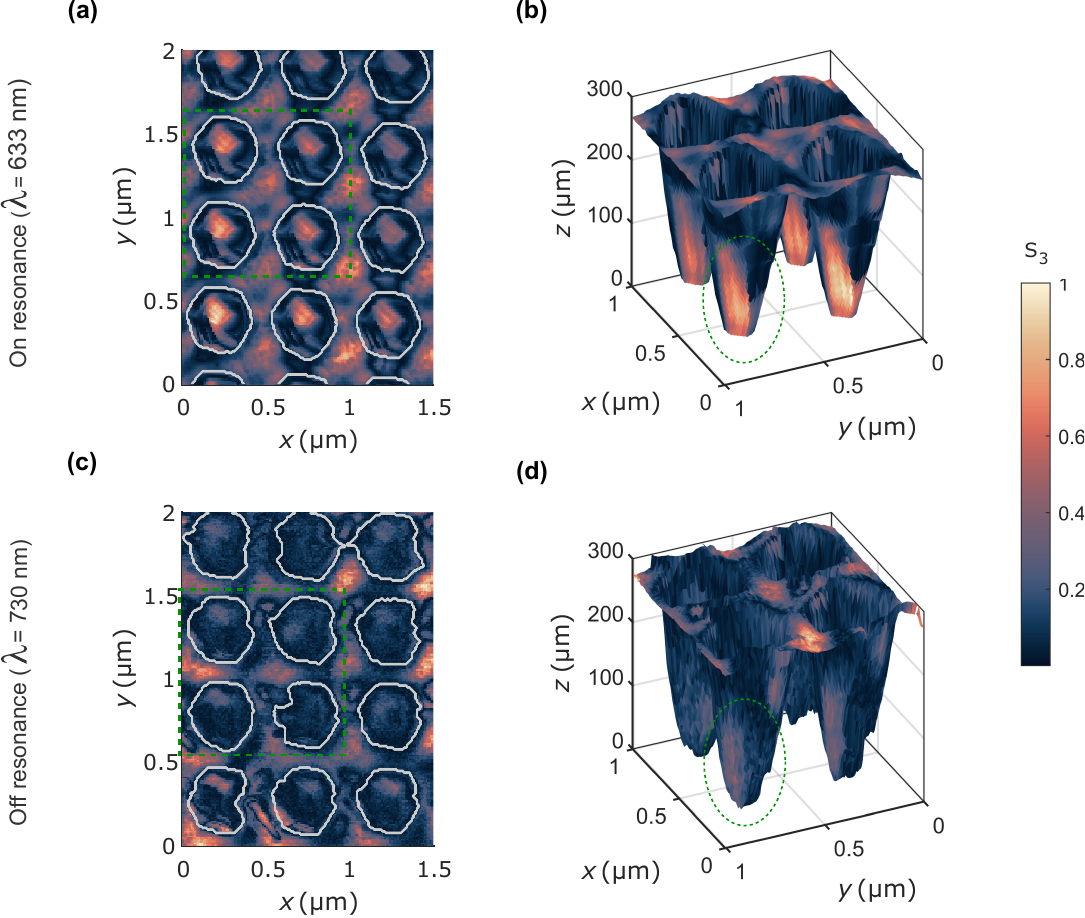}
    \caption{\textbf{Near-field measurements on and off resonance.} \textbf{(a)} Near-field intensity of the void array on resonance ($\lambda = \SI{633}{nm}$). The grey lines on the map show the position of the voids, based on the topography map. \textbf{(b)} Three-dimensional representation of the subset of voids indicated by the green dotted lines in (a). The green dotted circle highlights the field confinement inside the void. \textbf{(c)} Near-field intensity of the void array off resonance ($\lambda = \SI{730}{nm}$). (d) Three-dimensional representation of the subset of holes indicated by the green dotted lines in (d).
    The near-field scans are from the void array with $d = \SI{425}{nm}$.}
    \label{fig:SNOM}
\end{figure}

\begin{figure}[H]
    \centering
    \includegraphics[width=0.91\textwidth]{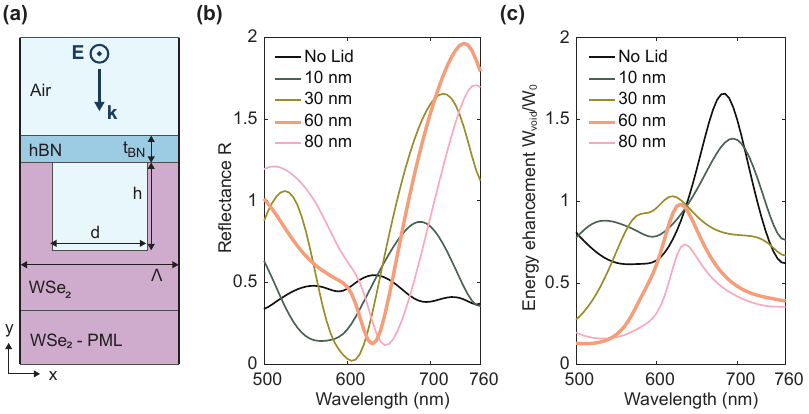}
    \caption{\textbf{hBN lid thickness.} \textbf{(a)} Schematic illustration of the 2D simulation geometry of a WSe$_2$ void with a hBN lid. The void dimensions are: $d = \SI{400}{nm}$, $h = \SI{450}{nm}$, and $\Lambda = \SI{500}{nm}$. The incident electric field has normal incidence and is polarized in the $z$-direction. \textbf{(b)} Reflectance and \textbf{(c)} energy enhancement spectra for increasing lid thicknesses. We find that thicknesses around \SI{60}{nm} are suitable for hBN lids.}
    \label{fig:BNLid_SI}
\end{figure}

\begin{figure}[H]
    \centering
    \includegraphics[width=0.91\textwidth]{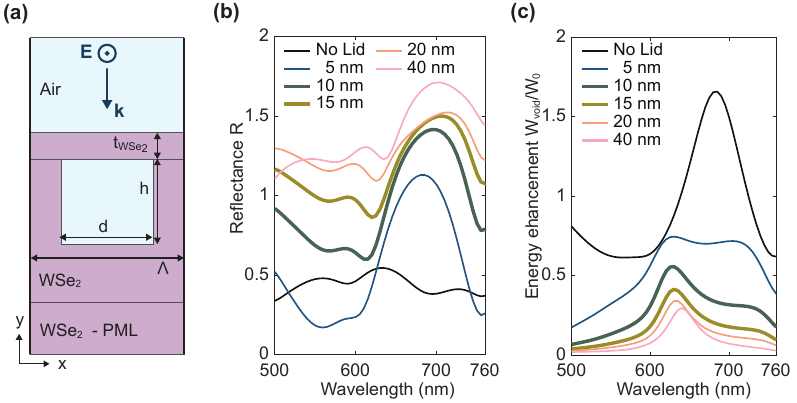}
    \caption{\textbf{WSe$_2$ lid thickness.} \textbf{(a)} Schematic illustration of the 2D simulation geometry for WSe$_2$ void with a WSe\textsubscript{2} lid. The void dimensions are: $d = \SI{400}{nm}$, $h = \SI{450}{nm}$, $\Lambda = \SI{500}{nm}$. \textbf{(b)} Reflectance and \textbf{(c)} energy spectra for increasing lid thicknesses. We find that thicknesses around 10-\SI{15}{nm} are suitable for WSe$_2$ lids.}
    \label{fig:WSe2Lid_SI}
\end{figure}

\begin{figure}[H]
    \centering
    \includegraphics{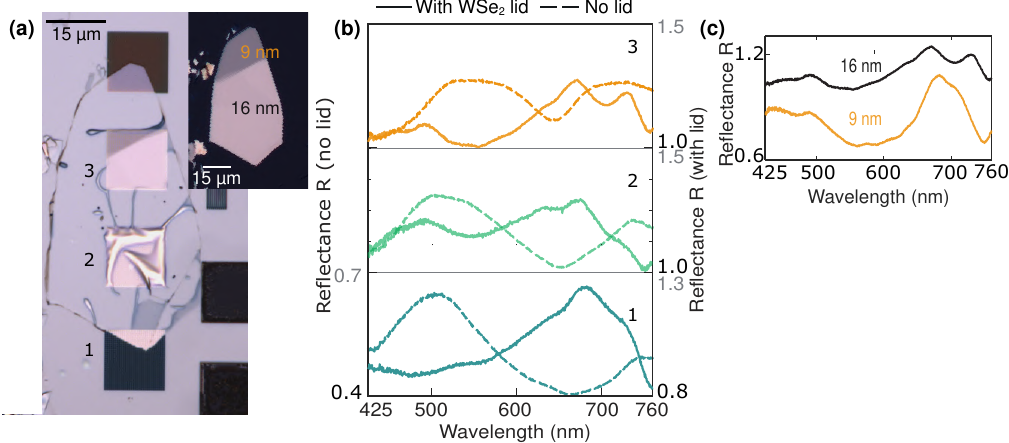}
    \caption{\textbf{Experimental encapsulation with WSe$_2$ lid.} \textbf{(a)} BF micrographs of voids fabricated on 422~nm thick flake and covered with a WSe$_2$ lid. Inset: The WSe$_2$ flake used as a lid consists of two regions with thicknesses of 9~nm and 16~nm.  \textbf{(b)} Experimental reflectance spectra from the voids in (a) with (solid) and without (dashed) the 16~nm thick WSe$_2$ lid. Left (right) axis shows the the reflectance of the open (encapsulated) voids. \textbf{(c)} Experimental reflectance spectra from void array 3 with the largest diameter and depth with a 9~nm and 16~nm thick WSe$_2$ lid.}
    \label{fig:WSe2 cap}
\end{figure}


\end{document}